\documentclass[aps,pra,twocolumn,groupedaddress,showpacs,floatfix]{revtex4}

\usepackage[dvips]{graphicx}
\usepackage{epsfig}
\usepackage{amssymb}
\usepackage{amsmath}
\usepackage{color}

\begin{document}


\title{Probing interactions between Rydberg atoms with large electric dipole moments in amplitude modulated electric fields}


\author{V. Zhelyazkova and S. D. Hogan}

\affiliation{Department of Physics and Astronomy, University College London, Gower Street, London WC1E 6BT, U.K.}


\date{\today}

\begin{abstract} Dipole-dipole interactions between helium atoms in Rydberg-Stark states with principal quantum number $n=53$ and approximately linear Stark energy shifts, resulting from induced electric dipole moments of approximately 7900~D, have been investigated experimentally. The experiments were performed in pulsed supersonic metastable helium beams, with particle number densities of up to $\sim10^9$~cm$^{-3}$. In the presence of amplitude-modulated, radio-frequency electric fields, changes in the spectral intensity distributions associated with the transitions to these states that are attributed to dipole-dipole interactions within the ensembles of excited atoms have been observed. The experimental results are in excellent agreement with calculations of the Rydberg energy level structure carried out using Floquet methods, and excitations shared by up to 4 atoms. The use of these Rydberg-Stark states as sensors for non-resonant broadband radio-frequency electrical noise is also discussed.
\end{abstract}


\maketitle

Atoms and molecules in Rydberg-Stark states with high principal quantum number, $n$, can possess very large induced electric dipole moments, $\mu$. For each value of $n$, the maximum dipole moment is $\mu_{\mathrm{max}}\simeq(3/2)n^2 e\,a_0$, where $e$ and $a_0$ are the electron charge and the Bohr radius, respectively~\cite{gallagher94a}. These dipole moments exceed 1000~D for $n>16$, and can be exploited to control the translational motion of gas-phase samples using inhomogeneous electric fields~\cite{wing80a,breeden81a,townsend01a,yamakita04a,vliegen04a,ko14a}. This has given rise to the realization of very efficient methods for acceleration, deceleration and electric trapping of atoms and molecules initially traveling in pulsed supersonic beams~\cite{vliegen06a,hogan08a,hogan09a,hogan12a,allmendinger13a,lancuba14a}. However, the large electric dipole moments of Rydberg-Stark states can also give rise to (i) strong electric dipole interactions within ensembles of atoms or molecules upon photoexcitation, or when confined in traps~\cite{reinhard07a,vogt07a}, and (ii) modifications to the energy level structure in the presence of radio-frequency (rf) electrical noise, or the effective time-dependent electric fields experienced by particles undergoing oscillatory motion in electrostatic traps~\cite{hogan08a}. 

We report here the results of experiments in which dipole-dipole interactions between helium (He) atoms in Rydberg-Stark states with electric dipole moments of $\simeq7900$~D have been studied in amplitude-modulated electric fields. Detailed experimental studies of the interactions between atoms in these states have not previously been carried out but are of importance for the interpretation and refinement of experiments with trapped atoms and molecules. The results presented are also essential for experiments directed toward the deterministic preparation of single atoms in circular Rydberg states using the crossed-field method~\cite{hyafil04a,delande88a,hare88a,anderson13a}, and for planned experiments in which cold Rydberg atoms are to be transported and trapped close to superconducting microwave circuits for applications in hybrid quantum information processing~\cite{hogan12b,carter13a,hermann14a}. In addition, the sensitivity of atoms in Rydberg-Stark states to weak rf electric fields, demonstrated here, highlights potential applications in broadband electrical noise sensing~\cite{holloway14a}. 

In experiments with atoms in Rydberg states that exhibit strong linear Stark energy shifts and large electric dipole moments, the interaction between pairs of atoms with dipole moments $\vec{\mu}_{1}$ and $\vec{\mu}_{2}$ , separated by a distance $R = |\vec{R}\,|$, corresponds to that of two interacting classical dipoles. The resulting interaction potential, $V_{\mathrm{dd}}$, is therefore:
\begin{eqnarray}
V_{\mathrm{dd}} &=& \frac{\mu_{1}\,\mu_{2}}{4\pi\epsilon_0 R^3}(1-3\cos^2\theta),\label{eq:dipdip}
\end{eqnarray}
where $\theta$ is the orientation of the position vector $\vec{R}$, and $\epsilon_0$ is the vacuum permittivity~\cite{comparat10a,reinhard07a}. This, effectively classical, dipole-dipole interaction differs from the resonant dipole-dipole interactions that arise from the large transition dipole moments associated with high Rydberg states~\cite{anderson98a,mourachko98a,vogt06a,saquet10}, and the, second-order, van der Waals interactions often exploited in Rydberg excitation blockade experiments and studies of many-body effects in ensembles of cold atoms~\cite{tong04a}. In the Rydberg-Stark states of He studied here, for which $\mu_1=\mu_2\simeq 7900$~D, Eq.~\ref{eq:dipdip} leads to an average dipole-dipole interaction energy of 10~MHz for a mean atom-atom separation of $\langle R\rangle \sim 7.8~\mu$m. This corresponds to a particle number density of $\rho\simeq \langle R\rangle^{-3}\simeq 2.1\times10^9$~cm$^{-3}$. In contrast to electric-field induced dipole blockade experiments reported previously~\cite{vogt07a}, the Rydberg-Stark states considered here are strongly $\ell$-mixed ($\ell$ is the orbital angular momentum quantum number of the Rydberg electron). It is this $\ell$-mixing that gives rise to the approximately linear Stark energy shifts, and the particularly large electric dipole moments upon which the experiments rely.
\begin{figure}[h!]
\includegraphics[width=0.40\textwidth]{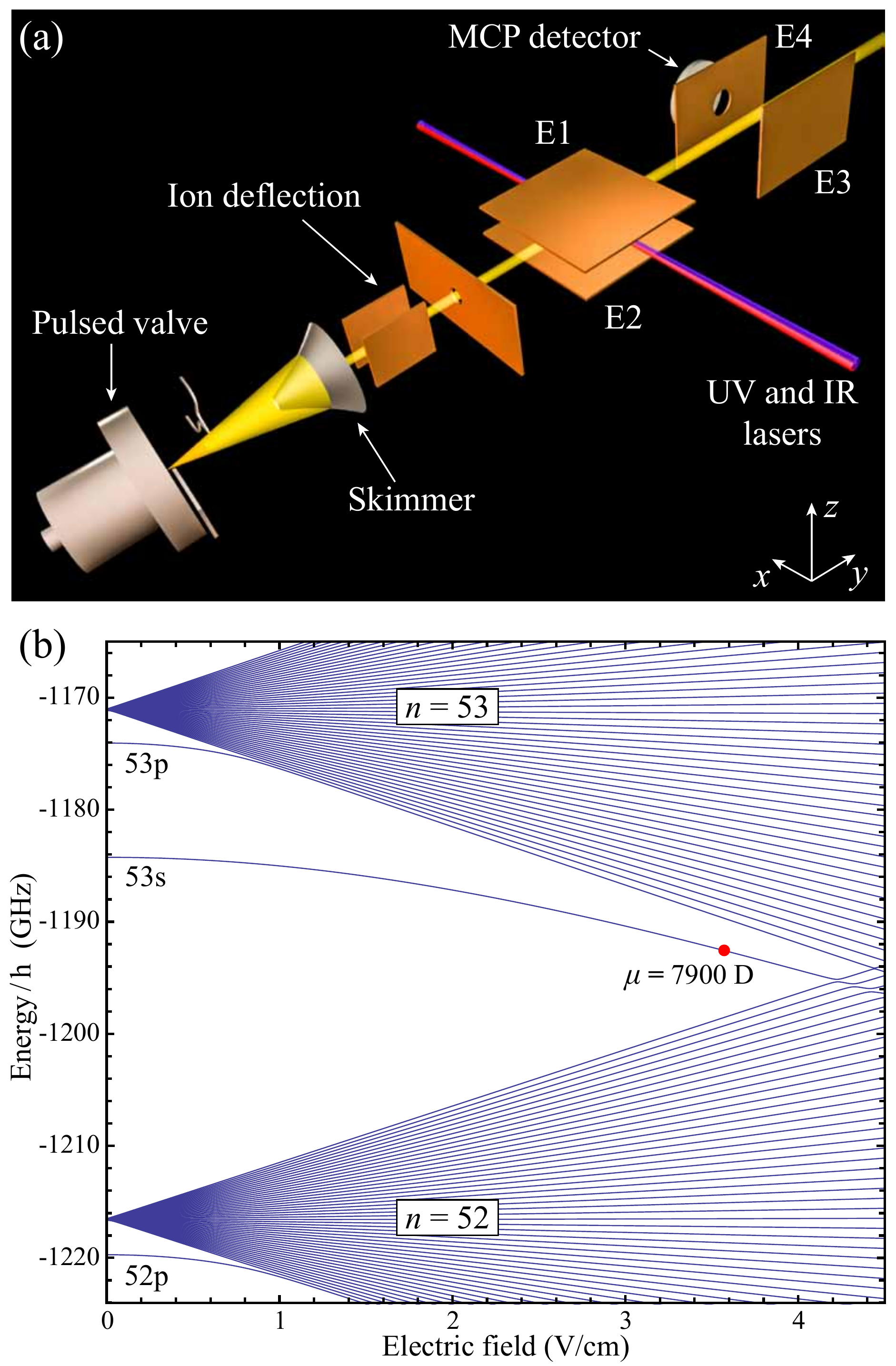}
\caption{\label{fig1}(Color online) (a) Schematic diagram of the experimental apparatus (not to scale). (b) Stark map of the triplet $n=52$ and $n=53$ Rydberg-Stark states in He. The red dot indicates the state excited in the electric fields of the experiment. In an electric field of 3.57~V\;cm$^{-1}$ this state has an electric dipole moment of $\simeq7900$~D.}
\end{figure}

A schematic diagram of the experimental apparatus is presented in Fig.~\ref{fig1}. A pulsed supersonic beam of metastable He atoms is generated in a dc electric discharge at the exit of a pulsed valve~\cite{halfmann00}. The stagnation pressure on the high pressure side of the valve is 6 bar, and the pulse duration is 200~$\mu$s. The resulting metastable He beam travels at 1950~m\,s$^{-1}$. After the atoms exit the source vacuum chamber through a copper skimmer (diameter 2 mm) they pass through an ion deflection zone where ions produced in the discharge are completely filtered out. The atoms then enter a region between two parallel 70 mm$\times$ 70 mm square metal plates, E1 and E2 in Fig.~\ref{fig1}(a), separated in the $z$ dimension by 8.4~mm. In this region the atomic beam is crossed by two copropagating laser beams which drive a resonant $1\mathrm{s}2\mathrm{s}\,^3\mathrm{S}_1\rightarrow1\mathrm{s}3\mathrm{p}\,^3\mathrm{P}_2\rightarrow1\mathrm{s}n\mathrm{s}/1\mathrm{s}n\mathrm{d}$ two-photon transition. The first photoexcitation step is driven by an ultraviolet (UV) laser operated at 388.975~nm which is stabilized using a He discharge cell~\cite{moron12}. Atoms in the $1\mathrm{s}3\mathrm{p}\,^3\mathrm{P}_2$ level are then excited to $1\mathrm{s}n\mathrm{s}/1\mathrm{s}n\mathrm{d}$ Rydberg states by an infrared (IR) laser operated at $\sim787$~nm. Both laser beams are linearly polarized parallel to the electric field axis ($z-$axis) and have Gaussian intensity profiles with full-widths-at-half-maxima (FWHM) in the $z$ and $y$ dimensions of 95 and 125~$\mu$m (204 and 80~$\mu$m), respectively, in the case of the UV (IR) radiation. The UV laser power is set to 35~$\mu$W (corresponding to an average intensity of $\sim$0.38~W\,cm$^{-2}$) to minimize effects of saturation. The power of the IR beam is varied in the range $4.6\,-\,230$~mW (with corresponding average intensities of $I_{\mathrm{IR}}=36\,-\,1794$~W\,cm$^{-2}$).

For all experiments described here, photoexcitation was carried out to the outermost high-field-seeking state in the $n=53$ Stark manifold, which evolves adiabatically to the $1\mathrm{s}53\mathrm{s}\,^3\mathrm{S}_1$ level in zero electric field. In an electric field $F_0=3.57$~V\,cm$^{-1}$ [red dot in Fig.~\ref{fig1}(b)] this state has $\sim56$\% s- and $\sim15$\% d-character, an approximately linear Stark shift, and an electric dipole moment of $\mu\simeq7900$~D. After photoexcitation the Rydberg atoms fly into the detection region of the apparatus where a pulsed potential of +1.4~kV is applied to the metal plate E3 in Fig.~\ref{fig1}(a), generating an electric field of $\sim378$~V\,cm$^{-1}$. This field ionizes the Rydberg atoms and accelerates the resulting He$^+$ ions through an aperture in E4, which is grounded, and onto a microchannel plate (MCP) detector.
\begin{figure}[!t]
\includegraphics*[width=0.5\textwidth,trim = 0mm 0mm 0mm 15mm]{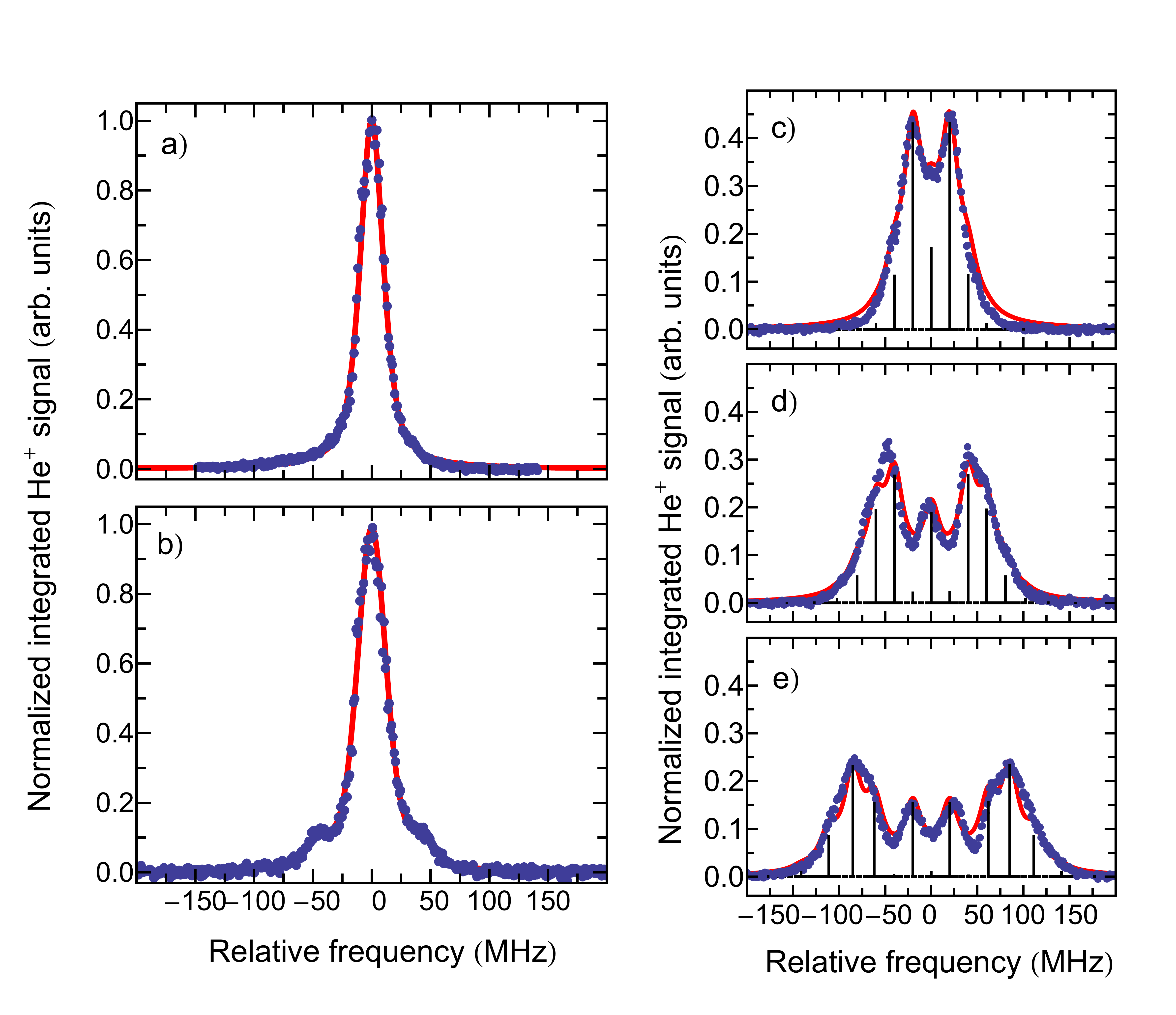}
\caption{\label{fig2} Spectra encompassing the transition to the outermost high-field-seeking $n=53$ Rydberg-Stark state. (a) Unperturbed spectrum, (b) as (a) but recorded in the presence of rf laboratory noise. (c)-(e) Spectra recorded in amplitude modulated electric fields (see Equation~\ref{elField}) for which $\omega_{\mathrm{rf}}=2\pi\times20$~MHz, and (c) $F_{\mathrm{osc}}=9.0$~mV\,cm$^{-1}$, (d) $F_{\mathrm{osc}}=18.0$~mV\,cm$^{-1}$ and (e) $F_{\mathrm{osc}}=27.0$~mV\,cm$^{-1}$. In each panel the blue points represent the experimental data, the black vertical bars correspond to the calculated Floquet sideband intensities, and the red curves represent a convolution of the calculated intensities with a Lorentzian function with a FWHM of 24~MHz. The spectra in each panel are normalized to the amplitude of the experimental data in (a).}
\end{figure}

Because the electric dipole moment of this Rydberg state is so large, it is very sensitive to dc and rf electric fields. An example of this can be seen in Fig.~\ref{fig2} which contains spectra encompassing the transition to this state when $F_0=3.57$~V\,cm$^{-1}$. In the absence of significant rf laboratory noise this transition exhibits a Lorentzian spectral profile [Fig.~\ref{fig2}(a)]. However, if the electrical noise emanating from the He rf discharge cell used to stabilize the UV laser is not shielded, the spectral profile of the transition changes, with the appearance of a broadened pedestal at low spectral intensities beneath the main Lorentzian feature [Fig.~\ref{fig2}(b)]. The origin of this lineshape can be understood by considering the effect of a weak rf amplitude modulation of the electric field on the Rydberg-Stark state. Such a field, $F(t)$, can be expressed as
\begin{equation}
F(t)=F_0+F_{\mathrm{osc}}\cos(\omega_{\mathrm{rf}}t),
\label{elField}
\end{equation}
where $F_{\mathrm{osc}}\ll F_0$ is the amplitude of the modulation which occurs at an angular frequency $\omega_{\mathrm{rf}}$. The oscillatory component of this field perturbs the time-independent Stark Hamiltonian and can be treated using Floquet methods~\cite{shirley65,zhang94,spellmeyer97,spellmeyer98,vanditzhuijzen09,yoshida12,tretyakov14}. This involves the construction of the Hamiltonian matrix in an $|n,\ell,m_{\ell},q\rangle$ basis, where $m_{\ell}$ is the azimuthal quantum number, and $q$ is an integer representing the number of rf photons associated with each Floquet sideband. The energies of the diagonal elements in this matrix were determined using the Rydberg formula with quantum defects $\delta_{\mathrm{s}}=0.2967$, $\delta_{\mathrm{p}}=0.0683$, $\delta_{\mathrm{d}}=0.0029$ and $\delta_{\mathrm{f}}=0.0004$ for the s, p, d, and f states, respectively~\cite{martin87a}. States for which $50\leq n\leq56$ were included in the matrix, each with $2q_{\mathrm{max}}+1$ Floquet sidebands, ranging in energy from $-q_{\mathrm{max}}\hbar\omega_{\mathrm{rf}}$ to $+q_{\mathrm{max}}\hbar\omega_{\mathrm{rf}}$ ($q_{\mathrm{max}}$ is the maximum value of $q$ considered). For all calculations presented convergence was achieved for $q_{\mathrm{max}}=8$. 

The dc electric field gives rise to off-diagonal matrix elements coupling states with $\Delta \ell=\pm 1$, and $\Delta q=0$, while the parallel rf field couples states with $\Delta\ell=\pm1$ and $\Delta q=0,\pm 1$~\cite{spellmeyer97,spellmeyer98}. The transition frequencies and spectral intensities were determined by calculating the eigenvalues and eigenvectors of the Hamiltonian matrix. The results of calculations carried out for $F_0=3.57$~V\,cm$^{-1}$, $\omega_{\mathrm{rf}}=2\pi\times20$~MHz, and a range of values of $F_{\mathrm{osc}}$ can be seen in Fig.~\ref{fig2}(c)-(e) (vertical bars), together with experimental data (blue dots) recorded for the same parameters. In these figures, the calculated spectral intensities have also been convoluted with Lorentzian functions with FWHM of $24$~MHz (red curves). The spectrum recorded for $F_{\mathrm{osc}}=0$ is that in Fig.~\ref{fig2}(a). For low-amplitude modulation, e.g., $F_{\mathrm{osc}}=9.0$~mV\,cm$^{-1}$ [Fig.~\ref{fig2}(c)], the unperturbed spectral feature is modified, with the intensity distributed over the low-$q$ rf sidebands. In this case, the $q=\pm1$ sidebands dominate. As $F_{\mathrm{osc}}$ is increased to 18.0~mV\,cm$^{-1}$ [Fig.~\ref{fig2}(d)], and 27.0~mV\,cm$^{-1}$ [Fig.~\ref{fig2}(e)], the spectral intensity is further displaced toward higher values of $q$. 

\begin{figure}[h]
\includegraphics[clip=,width=0.3\textwidth]{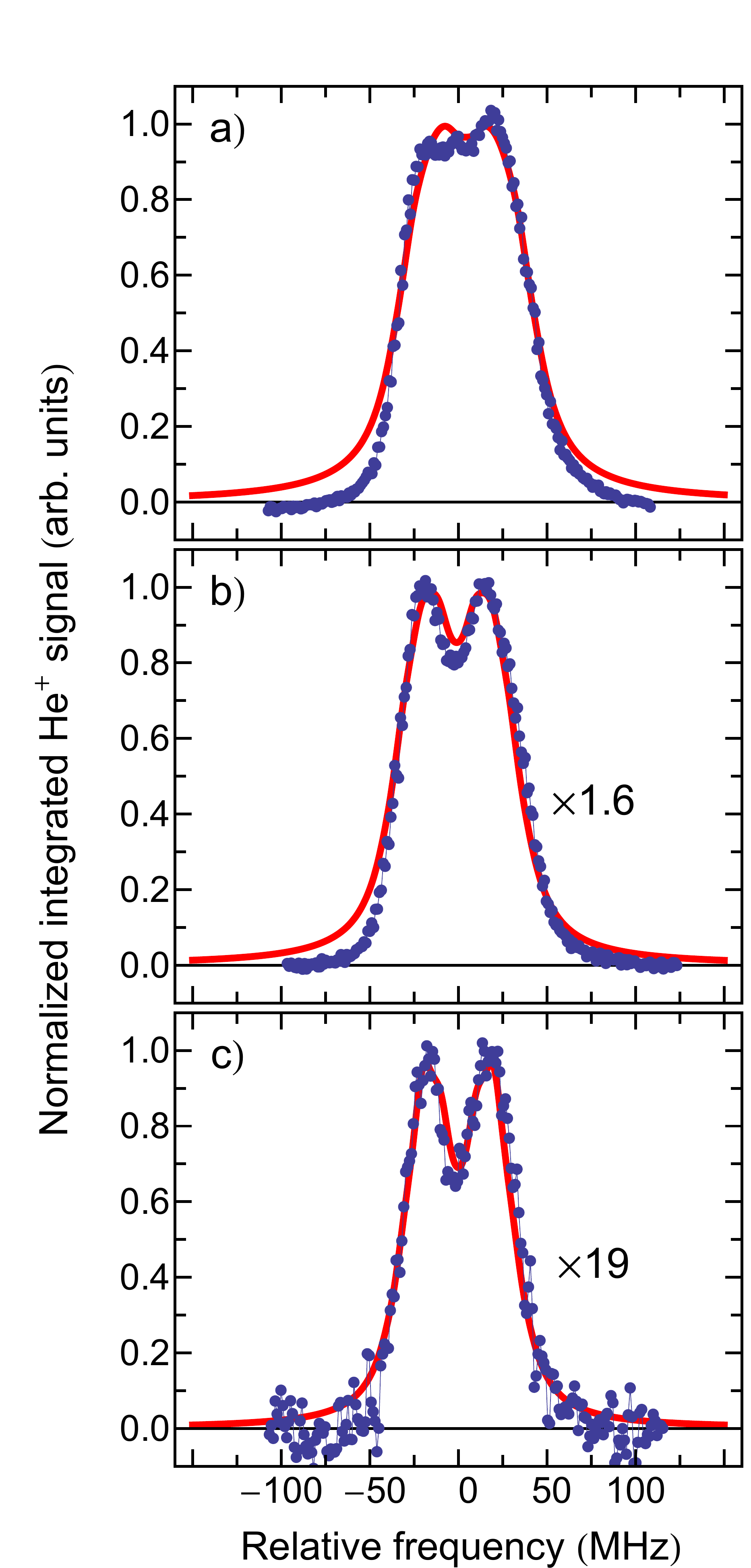}
\caption{\label{fig3} Experimentally recorded (blue dots) and calculated (red curves) spectra encompassing the transition to the outermost $n=53$ Rydberg-Stark state with $F_{\mathrm{osc}}=7.0$~mV\,cm$^{-1}$, $\omega_{\mathrm{rf}}=2\pi\times10$~MHz, and (a) $I_{\mathrm{IR}}=1794$~W\,cm$^{-2}$, (b) $I_{\mathrm{IR}}=718$~W\,cm$^{-2}$, and (c) $I_{\mathrm{IR}}=36$~W\,cm$^{-2}$. The calculations were performed for excitations shared by $N=4$ atoms (see text for details).}
\end{figure}

The agreement between the results of the calculations and the experimental data in Fig.~\ref{fig2}(c)-(e) indicates that the above approach can be extended to characterize ambient rf noise in the laboratory, such as that present when recording the spectrum in Fig.~\ref{fig2}(b). By comparing this data to a set of calculated spectra we find that the experimentally recorded spectral profile is characteristic of the presence of an oscillating field with $F_{\mathrm{osc}}=3.0$~mV\,cm$^{-1}$ and $\omega_{\mathrm{rf}}=2\pi\times43$~MHz. We conclude that the source of this noise is the second harmonic of the rf driving field of the He discharge cell used to stabilize the UV laser. In this rf frequency regime our measurements are sensitive to changes in $F_{\mathrm{osc}}$ of $\geq500~\mu$V\,cm$^{-1}$.

Inspired by electromagnetically induced transparency (EIT) studies of many-body effects in Rydberg ensembles~\cite{mohapatra07,weatherill08}, we have investigated how spectra recorded for selected values of $\omega_{\mathrm{rf}}$ and $F_{\mathrm{osc}}$ change depending upon the IR laser intensity and the particle number density (see Fig.~\ref{fig3}). These experiments were performed with $\omega_{\mathrm{rf}}=2\pi\times10$~MHz and $F_{\mathrm{osc}}=7.0$~mV\,cm$^{-1}$. Under these conditions and a low IR laser intensity ($I_{\mathrm{IR}}=36$~W\,cm$^{-2}$), the spectral intensity distribution exhibits a minimum at the unperturbed transition frequency arising from maximal suppression of the $q=0$ sideband, and intensity maxima in the $q=\pm1$ sidebands [Fig.~\ref{fig3}(c)]. As $I_{\mathrm{IR}}$ is increased to 718~W\,cm$^{-2}$ [Fig.~\ref{fig3}(b)] and 1794~W\,cm$^{-2}$ [Fig.~\ref{fig3}  (a)], the observed spectral intensity distributions change. Notably, in Fig.~\ref{fig3}(a), the minimum at the unperturbed transition frequency is absent and the transitions to the most intense sidebands are saturated.

To quantify the effects contributing to the changes in the experimental spectra in Fig.~\ref{fig3}, we calculate their dependence on $I_{\mathrm{IR}}$. In these calculations the electric dipole transition moments from the intermediate $1\mathrm{s}3\mathrm{p}\,^3\mathrm{P}_2$ level to each individual Floquet sideband were combined with the measured laser intensities and an unperturbed spectral linewidth of $\gamma=17$~MHz. This linewidth was determined from experimental data in Fig.~\ref{fig3}(a) where effects of saturation are negligible. Using this information, the saturation parameters $s_q=2\Omega_{s,q}^2/\gamma^2$ (where $\Omega_{s,q}$ is the single-atom Rabi frequency) were calculated for each Floquet sideband to determine the overall spectral intensity distribution function~\cite{metcalf99a}. We note that this procedure does not involve any free or fitted parameters. However, to obtain quantitative agreement between the results of the calculation and the experimental data, it was necessary to introduce a constant factor of 2$\pm0.25$ by which the single-atom Rabi frequencies were scaled. This factor is independent of laser intensity and suggests that excitations in the atomic beam are shared by multiple atoms coupled via dipole-dipole interactions. By making the assumption that the resulting many-body Rabi frequency is $\Omega_{N}=\sqrt{N}\,\Omega_{\mathrm{s}}$~\cite{stanojevic09,gaetan09,urban09}, where $N$ is the effective number of atoms sharing each excitation, best agreement between the experimental data and the calculations in Fig.~\ref{fig3} was obtained for $N=4\pm1$ atoms. The deviation of the wings of the calculated spectral profiles from the experimental data in Fig.~\ref{fig3}(a) and (b) is attributed to effects of the Gaussian spatial intensity distributions of the laser beams, and variations in the inter-particle spacing within the atomic beam. Only the spatially averaged IR laser intensities were considered in the calculations. 

\begin{figure}[t]
\centering
\includegraphics[width=0.486\textwidth]{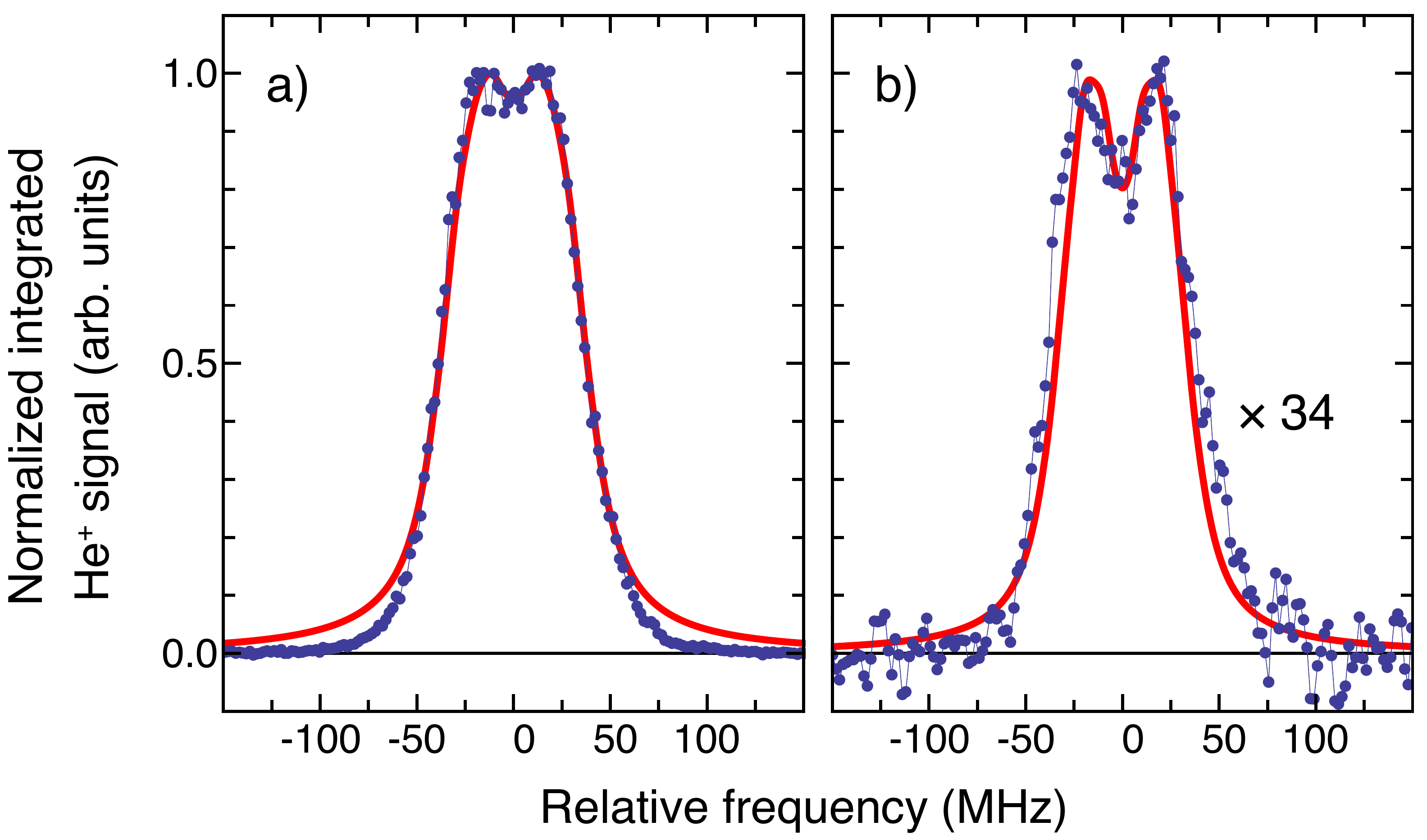}
\caption{\label{fig4} Experimentally recorded (blue dots) and calculated (red curves) spectral intensity distributions in (a) high, and (b) low particle number density regimes, for $I_{\mathrm{IR}}=1615$~W\,cm$^{-2}$. The calculations in (a) and (b) were performed for $N=4$ and $N=1$, respectively. At low density the unsaturated Floquet sideband intensity distribution, characterized by the minimum at the unperturbed transition frequency, is recovered with a contrast of $\sim80$\%.}
\end{figure}

To confirm that the changes observed in the spectra in Fig.~\ref{fig3} do indeed arise as a result of many-body interactions, it is necessary to demonstrate that by reducing the atom number density the unsaturated Floquet sideband intensity distribution is recovered. This was achieved by adjusting the timings in the experiment so that the low density trailing component of the pulsed supersonic beam was probed. Spectra recorded for high and low atom number densities but a constant value of $I_{\mathrm{IR}}=1615$~W\,cm$^{-2}$) are displayed in Fig.~\ref{fig4} (blue dots). These measurements were performed for the same values of $\omega_{\mathrm{osc}}$ and $F_{\mathrm{osc}}$ as in Fig.~\ref{fig3}, and approximately equal particle number density, in the case of Fig.~\ref{fig4}(a). Comparison of the experimental data (blue dots) recorded at high density [Fig.~\ref{fig4}(a)] with the calculated spectra indicates that, as in Fig.~\ref{fig3}, each excitation is shared by $N\simeq4$ atoms. In the low density regime, in which the amplitude of the signal is reduced by a factor of 34 [Fig.~\ref{fig4}(b)], the unsaturated Floquet sideband intensity distribution is recovered with its characteristic minimum clearly visible at the unperturbed transition frequency. The calculations indicate that in the low density case excitations are associated with single atoms, i.e., $N=1$. In this case the dip at the unperturbed transition frequency is approximately 80\% of the intensity of the spectral maxima. We define this as the contrast which, in the experimental spectrum in Fig.~\ref{fig4}(b), is $80\pm5$\% and in very good agreement with the results of the calculations. In the higher density case the calculated spectral intensity distributions for $N=3,4$ and 5 result in a contrast of 92, 96 and 98\%, respectively. The contrast in the experimental data in Fig.~\ref{fig4}(a) is 95$\pm2$\% which is in closest agreement with the calculation for which $N=4$. 

Averaging over the angular dependence of the dipole-dipole interaction potential, Eq.~\ref{eq:dipdip}, results in a blockade radius of 8.2~$\mu$m for the unsaturated spectral half-width-at-half-maximum of $\gamma/2=8.5$~MHz. If 4 atoms are considered to be located within a sphere of this radius a particle number density of $\rho=1.8\times10^9$~cm$^{-3}$ results. This density agrees well with those previously achieved in similar pulsed supersonic atomic beams~\cite{saquet10,halfmann00}. 

The results of the experiments and calculations reported here demonstrate the sensitivity of Rydberg-Stark states with large electric dipole moments to non-resonant rf electric fields, and to dipole-dipole interactions. The detection of changes in the spectral intensity distribution associated with Rydberg excitation in the presence of a low-frequency amplitude modulated electric fields, represents a direct spectroscopic observable which can be employed to identify many-body contributions to the Rydberg photoexcitation process, without the need to directly count the number of atoms contributing to each excitation. In addition, by acting as microscopic rf antennas, Rydberg atoms can be exploited as broadband probes of electrical noise. They are ideally suited to the characterization of rf noise at vacuum--solid-state interfaces of importance, for example, in hybrid approaches to quantum information processing~\cite{hogan12b,hermann14a}, and in chip-based ion traps~\cite{turchette00}. Combining the techniques presented here with Rydberg atom detection by EIT~\cite{mohapatra07} could, in the future, permit direct optical readout of such noise.

The observed effects of electric dipole interactions on the saturation of optical spectra encompassing transitions to Rydberg states with electric dipole moments of $\simeq7900$~D demonstrates the possibility of investigating many-body physics in pulsed supersonic beams with particle number densities of $\sim2\times10^{9}$~cm$^{-3}$. Of particular interest will be studies of many-body contributions in atom-molecule scattering at low collision energies. Elucidating the role, as reported here, played by these interactions within ensembles of Rydberg atoms and molecules is also crucial for the refinement of experiments directed toward investigations of the decay mechanisms of electrically trapped samples~\cite{seiler11}.

\begin{acknowledgments}
This work was supported financially by the Department of Physics and Astronomy and the Faculty of Mathematical and Physical Sciences at University College London, and the Engineering and Physical Sciences Research Council under Grant No.~EP/L019620/1.
\end{acknowledgments}

\end{document}